\def\year{2020}\relax
\documentclass[letterpaper]{article} 
\usepackage{aaai20}  
\usepackage{times}  
\usepackage{helvet} 
\usepackage{courier}  
\usepackage[hyphens]{url}  
\usepackage{graphicx} 
\usepackage{soul}
\usepackage{cuted}
\usepackage{xcolor}
\usepackage{textcomp}
\usepackage{capt-of}
\usepackage{gensymb}
\usepackage{graphicx}  
\nocopyright 
\title{Interactive Task and Concept Learning from Natural Language Instructions and GUI Demonstrations\thanks{To be presented at the AAAI-20 Workshop on Intelligent Process Automation (IPA-20). An earlier version of this paper was published and presented at ACM UIST 2019.}}
\author{\Large \textbf{Toby Jia-Jun Li\textsuperscript{\rm 1}, Marissa Radensky\textsuperscript{\rm 2}, Justin Jia\textsuperscript{\rm 1},} \\ \Large \textbf{Kirielle Singarajah\textsuperscript{\rm 1}, Tom M. Mitchell\textsuperscript{\rm 1}, Brad A. Myers\textsuperscript{\rm 1}}\\ 
\textsuperscript{\rm 1}Carnegie Mellon University, Pittsburgh, PA\\ 
\textsuperscript{\rm 2}Amherst College, Amherst, MA\\ 
}

\begin{document}

\maketitle

\begin{abstract}
Natural language programming is a promising approach to enable end users to instruct new tasks for intelligent agents. However, our formative study found that end users would often use unclear, ambiguous or vague concepts when naturally instructing tasks in natural language, especially when specifying conditionals. Existing systems have limited support for letting the user teach agents new concepts or explaining unclear concepts. In this paper, we describe a new multi-modal domain-independent approach that combines natural language programming and programming-by-demonstration to allow users to first naturally describe tasks and associated conditions at a high level, and then collaborate with the agent to recursively resolve any ambiguities or vagueness through conversations and demonstrations. Users can also define new procedures and concepts by demonstrating and referring to contents within GUIs of existing mobile apps. We demonstrate this approach in \textsc{Pumice}, an end-user programmable agent that implements this approach. A lab study with 10 users showed its usability.
\end{abstract}

\section{Introduction}
The goal of end user development (EUD) is to empower users with little or no programming expertise to program~\cite{Paterno:2017:NPE:3152674}. Among many EUD applications, a particularly useful one would be task automation, through which users program intelligent agents to perform tasks on their behalf~\cite{maes_agents_1994}. To support such EUD activities, a major challenge is to help non-programmers to specify conditional structures in programs.  Many common tasks involve conditional structures, yet they are difficult for non-programmers to correctly specify using existing EUD techniques due to the great distance between how end users think about the conditional structures, and how they are represented in programming languages~\cite{pane_studying_2001}. 

According to Green and Petre's cognitive dimensions of notations~\cite{green_usability_1996}, the closer the programming world is to the problem world, the easier the problem-solving ought to be. This \textit{closeness of mapping} is usually low in conventional and EUD programming languages, as they require users to think about their tasks very differently from how they would think about them in familiar contexts~\cite{pane_studying_2001}, making programming particularly difficult for end users who are not familiar with programming languages and  ``computational thinking''~\cite{wing_computational_2006}. To address this issue, the concept of \textit{natural programming}~\cite{myers_making_2017,myers_natural_2004} has been proposed to create techniques and tools that match more closely the ways users think.

\textit{Natural \underline{language} programming} is a promising technique for bridging the gap between user mental models of tasks and programming languages~\cite{mihalcea_nlp_2006}. It should have a low learning barrier for end users, under the assumption that the majority of end users can already communicate procedures and structures for familiar tasks through natural language conversations~\cite{lieberman_feasibility_2006,pane_studying_2001}. Speech is also a natural input modality for users to describe desired program behaviors~\cite{oviatt_ten_1999}. However, existing natural language programming systems are not adequate for supporting end user task automation in domain-general tasks. Some prior systems (e.g., ~\cite{price_naturaljava:_2000}) directly translate user instructions in natural language into conventional programming languages like Java. This approach requires users to use unambiguous language with fixed structures similar to those in conventional programming languages. Therefore, it does not match the user's existing mental model of tasks, imposing significant learning barriers and high cognitive demands on end users. 

Other natural language programming approaches (e.g.,~\cite{azaria_instructable_2016,fast_iris:_2018,kate_learning_2005,srivastava_joint_2017}) restricted the problem space to specific task domains, so that they could constrain the space and the complexity of target program statements in order to enable the understanding of flexible user utterances. Such restrictions are due to the limited capabilities of existing natural language understanding techniques -- they do not yet support robust understanding of utterances across diverse domains without extensive training data and structured prior knowledge within each domain.

Another difficult problem in natural language programming is supporting the instruction of \textit{concepts}. In our study (details below in the Formative Study section), we found that end users often refer to ambiguous or vague concepts (e.g., \textbf{cold} weather, \textbf{heavy} traffic) when naturally instructing a task. Moreover, even if a concept may seem clear to a human, an agent may still not understand it due to the limitations in its natural language understanding techniques and pre-defined ontology. 

In this paper, we address the research challenge of enabling end users to augment domain-independent task automation scripts with conditional structures and new concepts through a combination of natural language programming and programming by demonstration (PBD). To support programming for tasks in diverse domains, we leverage the graphical user interfaces (GUIs) of existing third-party mobile apps as a medium, where procedural actions are represented as sequences of GUI operations, and declarative concepts can be represented through references to GUI contents. This approach supports EUD for a wide range of tasks, provided that these tasks can be performed with one or more existing third-party mobile apps. 

We took a \textit{user-centered design} approach, first studying how end users naturally describe tasks with conditionals in natural language in the context of mobile apps, and what types of tasks they are interested in automating. Based on insights from this study, we designed and implemented an end-user-programmable conversational agent named \textsc{Pumice}\footnote{\textsc{Pumice} is a type of volcanic rock. It is also an acronym for \textbf{P}rogramming in a \textbf{U}ser-friendly \textbf{M}ultimodal \textbf{I}nterface through \textbf{C}onversations and \textbf{E}xamples}~\cite{li_pumice:_2019} that allows end users to program tasks with flexible conditional structures and new concepts across diverse domains through spoken natural language instructions and demonstrations.

\textsc{Pumice} extends our previous \textsc{Sugilite}~\cite{li_sugilite:_2017} system. A key novel aspect of \textsc{Pumice}'s design is that it allows users to first describe the desired program behaviors and conditional structures naturally in natural language at a high level, and then collaborate with an intelligent agent through multi-turn conversations to explain and to define any ambiguities, concepts and procedures in the initial description as needed in a top-down fashion. Users can explain new concepts by referring to either previously defined concepts, or to the contents of the GUIs of third-party mobile apps. Users can also define new procedures by demonstrating using third-party apps~\cite{li_sugilite:_2017}. Such an approach facilitates effective program reuse in automation authoring, and provides support for a wide range of application domains, which are two major challenges in prior EUD systems. The results from the motivating study suggest that this paradigm is not only feasible, but also natural for end users, which was supported by our summative lab usability study. 

We build upon recent advances in natural language processing (NLP) to allow \textsc{Pumice}'s semantic parser to learn from users' flexible verbal expressions when describing desired program behaviors. Through \textsc{Pumice}'s mixed-initiative conversations with users, an underlying persistent knowledge graph is dynamically updated with new procedural (i.e., actions) and declarative (i.e., concepts and facts) knowledge introduced by users, allowing the semantic parser to improve its understanding of user utterances over time. This structure also allows for effective reuse of user-defined procedures and concepts at a fine granularity, reducing user effort in EUD.

\textsc{Pumice} presents a multi-modal interface, through which users interact with the system using a combination of demonstrations, pointing, and spoken commands. Users may use any modality that they choose, so they can leverage their prior experience to minimize necessary training~\cite{maclellan_framework_2018,laird2017interactive}. This interface also provides users with guidance through a mix of visual aids and verbal directions through various stages in the process to help users overcome common challenges and pitfalls identified in the formative study, such as the omission of else statements, the difficulty in finding correct GUI objects for defining new concepts, and the confusion in specifying proper data descriptions for target GUI objects. A summative lab usability study with 10 participants showed that users with little or no prior programming expertise could use \textsc{Pumice} to program automation scripts for 4 tasks derived from real-world scenarios. Participants also found \textsc{Pumice} easy and natural to use.

This paper presents the following three primary contributions:

\begin{enumerate}
	
	\item A formative study showing the characteristics of end users' natural language instructions for tasks with conditional structures in the context of mobile apps. 
	
	\item A multi-modal conversational approach for the EUD of task automation motivated by the aforementioned formative study, with the following major advantages:
	
	\begin{enumerate}
		\item The top-down conversational structure allows users to \textbf{naturally} start with describing the task and its conditionals at a high-level, and then recursively clarify ambiguities, explain unknown concepts and define new procedures through conversations.
		
		\item The agent learns new procedural and declarative knowledge through explicit instructions from users, and stores them in a persistent knowledge graph, facilitating effective \textbf{reusability and generalizability} of learned knowledge.
		
		\item The agent learns concepts and procedures in \textbf{various task domains} while having a \textbf{low learning barrier} through its \textbf{multi-modal approach} that supports references and demonstrations using the contents of third-party apps' GUIs.

	\end{enumerate}
	
	\item The \textsc{Pumice} system: an implementation of this approach, along with a user study evaluating its usability.

\end{enumerate}

\section{Background and Related Work}
This research builds upon prior work from many different sub-disciplines across human-computer interaction (HCI), software engineering (SE), and natural language processing (NLP). In this section, we focus on related work on three topics: (1) natural language programming; (2) programming by demonstration; and (3) the multi-modal approach that combines natural language inputs with demonstrations.

\subsection{Natural Language Programming}
\textsc{Pumice} uses natural language as the primary modality for users to program task automation. The idea of using natural language inputs for programming has been explored for decades~\cite{ballard_programming_1979,biermann_natural_1983}. In the NLP and AI communities, this idea is also known as learning by instruction~\cite{azaria_instructable_2016,lieberman_instructible_1996}. 

The foremost challenge in supporting natural language programming is dealing with the inherent ambiguities and vagueness in natural language~\cite{vadas2005programming}. To address this challenge, one prior approach was to constrain the structures and expressions in the user's language to similar formulations of conventional programming languages (e.g.,~\cite{ballard_programming_1979,price_naturaljava:_2000}), so that user inputs can be directly translated into programming statements. This approach is not adequate for EUD, as it has a high learning barrier for users without programming expertise.\looseness=-1

Another approach for handling ambiguities and vagueness in natural language inputs is to seek user clarification through conversations. For example, Iris~\cite{fast_iris:_2018} asks follow-up questions and presents possible options through conversations when the initial user input is incomplete or unclear. This approach lowers the learning barrier for end users, as it does not require them to clearly define everything up front. It also allows users to form complex commands by combining multiple natural language instructions in conversational turns under the guidance of the system. \textsc{Pumice} also adopts the use of multi-turn conversations as a key strategy in handling ambiguities and vagueness in user inputs. However, a key difference between \textsc{Pumice} and other conversational instructable agents is that \textsc{Pumice} is domain-independent. All conversational instructable agents need to map the user's inputs onto existing concepts, procedures and system functionalities supported by the agent, and to have natural language understanding mechanisms and training data in each task domain. Because of this constraint, existing agents limit their supported tasks to one or a few pre-defined domains, such as data science~\cite{fast_iris:_2018}, email processing~\cite{azaria_instructable_2016,srivastava_joint_2017}, or database queries~\cite{kate_learning_2005}. 

\textsc{Pumice} supports learning concepts and procedures from existing third-party mobile apps regardless of the task domains. End users can create new concepts with \textsc{Pumice} by referencing relevant information shown in app GUIs, and define new procedures by demonstrating with existing apps. This approach allows \textsc{Pumice} to support a wide range of tasks from diverse domains as long as the corresponding mobile apps are available. This approach also has a low learning barrier because end users are already familiar with the capabilities of mobile apps and how to use them. In comparison, with prior instructable agents, it is often unclear what concepts, procedures and functionalities already exist to be used as ``building blocks'' for developing new ones.  

\subsection{Programming by Demonstration}
\textsc{Pumice} uses the programming by demonstration (PBD) technique to enable end users to define concepts by referring to the contents of GUIs of third-party mobile apps, and teach new procedures through demonstrations with those apps. PBD is a natural way of supporting EUD with a low learning barrier~\cite{cypher_watch_1993,lieberman_your_2001}. Many domain-specific PBD tools have been developed in the past in various domains, such as text editing (e.g.,~\cite{Lau:2001:LRT:369505.369519}), photo editing (e.g.,~\cite{grabler_generating_2009}), web scraping (e.g.,~\cite{Chasins:2018:RSD:3242587.3242661}), smart home control (e.g.,~\cite{li_programming_2017}) and robot control (e.g.,~\cite{argall_survey_2009}).

\textsc{Pumice} supports domain-independent PBD by using the GUIs of third-party apps for task automation and data extraction. Similar approaches have also been used in prior systems. For example, \textsc{Sugilite}~\cite{li_sugilite:_2017}, \textsc{Kite}~\cite{li_kite:_2018} and \textsc{Appinite}~\cite{li_appinite:_2018} use mobile app GUIs, CoScripter~\cite{leshed_coscripter:_2008}, d.mix~\cite{Hartmann:2007:PSR:1294211.1294254}, Vegemite~\cite{Lin:2009:EPM:1502650.1502667} and \textsc{Plow}~\cite{allen_plow:_2007} use web interfaces, and \textsc{Hilc}~\cite{Intharah:2019:HDP:3320251.3234508} and Sikuli~\cite{yeh_sikuli:_2009} use desktop GUIs. Compared to those, \textsc{Pumice} is the only one that can learn concepts as generalized knowledge, and the only one that supports creating conditionals from natural language instructions. Sikuli~\cite{yeh_sikuli:_2009} allows users to create conditionals in a scripting language, which is not suitable for end users without programming expertise.

\subsection{The Multi-Modal Approach}
A central challenge in PBD is generalization. A PBD agent should go beyond literal record-and-replay macros, and be able to perform similar tasks in new contexts~\cite{cypher_watch_1993,lieberman_your_2001}. This challenge is also a part of the program synthesis problem. An effective way of addressing this challenge is through multi-modal interaction~\cite{oviatt_ten_1999}. Demonstrations can clearly communicate \textit{what} the user does, but not \textit{why} the user does this and \textit{how} the user wants to do this in different contexts. On the other hand, natural language instructions can often reflect the user's underlying intent (\textit{why}) and preferences (\textit{how}), but they are usually ambiguous or unclear. This is where grounding natural language instructions with concrete GUI demonstrations can help.

This \textit{mutual disambiguation} approach~\cite{oviatt_mutual_1999} in multi-modal interaction has been proposed and used in many previous systems. This approach leverages \textit{repetition} in a different modality for mediation~\cite{mankoff2000oops}. Particularly for PBD generalization, \textsc{Sugilite}~\cite{li_sugilite:_2017} and \textsc{Plow}~\cite{allen_plow:_2007} use natural language inputs to identify parameterization in demonstrations, \textsc{Appinite}~\cite{li_appinite:_2018} uses natural language explanations of intents to resolve the ``data description''~\cite{cypher_watch_1993} for demonstrated actions, \textsc{Mars}~\cite{chen2019maximal} similarly used multi-layer specifications including the textual descriptions of user intents to disambiguate program synthesis results from user-provided input-output examples in the data science domain.

\textsc{Pumice} builds upon this prior work, and extends the multi-modal approach to support learning concepts involved in demonstrated tasks. The learned concepts can also be generalized to new task domains, as described in later sections. The prior multi-modal PBD systems also use demonstration as the main modality. In comparison, \textsc{Pumice} uses natural language conversation as the main modality, and uses demonstration for grounding unknown concepts, values, and procedures after they have been broken down and explained in conversations.

\section{Formative Study}
We took a \textit{user-centered} approach~\cite{myers_programmers_2016} for designing a natural end-user development system~\cite{myers_making_2017,radensky2018end}. We first studied how end users naturally communicate tasks with declarative concepts and control structures in natural language for various tasks in the mobile app context through a formative study on Amazon Mechanical Turk with 58 participants (41 of which are non-programmers; 38 men, 19 women, 1 non-binary person).\looseness=-1 

Each participant was presented with a graphical description of an everyday task for a conversational agent to complete in the context of mobile apps. All tasks had distinct conditions for a given task so that each task should be performed differently under different conditions, such as playing different genres of music based on the time of the day. Each participant was assigned to one of 9 tasks. To avoid biasing the language used in the responses, we used the Natural Programming Elicitation method~\cite{myers_programmers_2016} by showing graphical representations of the tasks with limited text in the prompts. Participants were asked how they would verbally instruct the agent to perform the tasks, so that the system could understand the differences among the conditions and what to do in each condition. Each participant was first trained using an example scenario and the corresponding example verbal instructions.

To study whether having mobile app GUIs can affect users' verbal instructions, we  randomly assigned participants into one of two groups. For the experimental group, participants instructed agents to perform the tasks while looking at relevant app GUIs. Each participant was presented with a mobile app screenshot with arrows pointing to the screen component  that contained the information pertinent to the task condition. Participants in the control group were not shown app GUIs. At the end of each study session, we also asked the participants to come up with another task scenario of their own where an agent should perform differently in different conditions.  

The participants' responses were analyzed by two independent coders using open coding~\cite{strauss_basics_1990}. The inter-rater agreement \cite{cohen_coefficient_1960} was $\kappa=0.87$, suggesting good agreement. 19\% of responses were excluded from the analysis for quality control due to the lack of efforts in the responses, question misunderstandings, and blank responses. 

We report the most relevant findings which motivated the design of \textsc{Pumice} next.

\subsection{App GUI Grounding Reduces Unclear Concept Usage}
We analyzed whether each user's verbal instruction for the task provided a clear definition of the conditions in the task. In the control group (instructing without seeing app screenshots), 33\% of the participants used ambiguous, unclear or vague concepts in the instructions, such as ``\textit{If it is daytime, play upbeat music...}'' which is ambiguous as to when the user considers it to be ``daytime.'' This is despite the fact that the example instructions they saw had clearly defined conditions. 

Interestingly, for the experimental group, where each participant was provided an app screenshot displaying specific information relevant to the task's condition, fewer  participants (9\%) used ambiguous or vague concepts (this difference is statistically significant with \textit{p} < 0.05), while the rest clearly defined the condition (e.g., ``\textit{If the current time is before 7 pm...}''). These results suggest that end users naturally use ambiguous and vague concepts when verbally instructing task logic, but showing users relevant mobile app GUIs with concrete instances of the values can help them ground the concepts, leading to fewer ambiguities and vagueness in their descriptions. The implication is that a potentially effective approach to avoiding unclear utterances for agents is to guide users to explain them in the context of app GUIs.

\subsection{Unmet User Expectation of Common Sense Reasoning}
We observed that participants often expected and assumed the agent to have the capability of understanding and reasoning with common sense knowledge when instructing tasks. For example, one user said, ``\textit{if the day is a weekend}''. The agent would therefore need to understand the concept of ``weekend'' (i.e., how to know today's day of the week, and what days count as ``weekend'') to resolve this condition. Similarly when a user talked about ``sunset time'', he expected the agent to know what it meant, and how to find out its value. 

However, the capability for common sense knowledge and reasoning is very limited in current agents, especially across many diverse domains, due to the spotty coverage and unreliable inference of existing common sense knowledge systems. Managing user expectation and communicating the agent's capability is also a long-standing unsolved challenge in building interactive intelligent systems~\cite{lieberman_beating_2004}. A feasible workaround is to enable the agent to ask users to ground new concepts to existing contents in apps when they come up, and to build up knowledge of concepts over time through its interaction with users.

\subsection{Frequent Omission of Else Statements}
In the study, despite all provided example responses containing else statements, 18\% of the 39 descriptions from users omitted an else statement when it was expected. ``\textit{Play upbeat music until 8pm every day,}'' for instance, may imply that the user desires an alternative genre of music to be played at other times. Furthermore, 33\% omitted an else statement when a person would be expected to infer an else statement, such as: ``\textit{If a public transportation access point is more than half a mile away, then request an Uber},'' which implies using public transportation otherwise. This might be a result of the user's expectation of common sense reasoning capabilities. The user omits what they expect the agent can infer to avoid prolixity, similar to patterns in human-human conversations~\cite{grice1975logic}.

These findings suggest that end users will often omit appropriate else statements in their natural language instructions for conditionals. Therefore, the agent should proactively ask users about alternative situations in conditionals when appropriate.

\begin{figure*}
	\centering
	\includegraphics[width=0.95\textwidth]{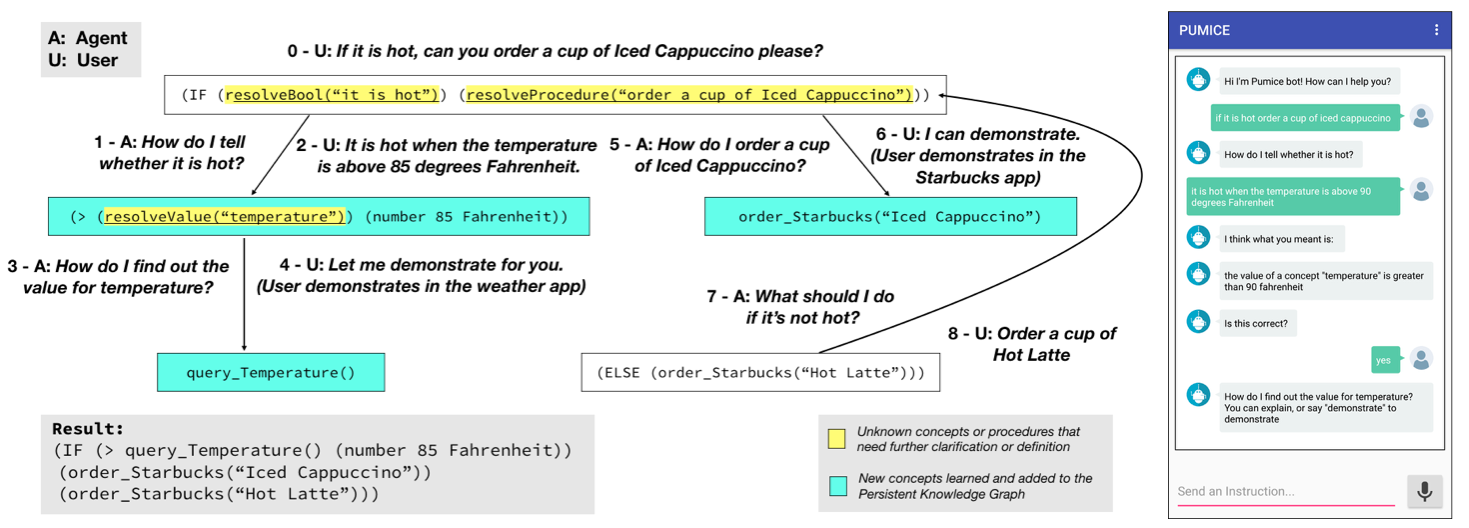}
	\vspace{-0.4cm}
	\caption{Example structure of how \textsc{Pumice} learns the concepts and procedures in the command ``If it's hot, order a cup of Iced Cappuccino.'' The numbers indicate the order of utterances. The screenshot on the right shows the conversational interface of \textsc{Pumice}. In this interactive parsing process, the agent learns how to query the current temperature, how to order any kind of drink from Starbucks, and the generalized concept of ``hot'' as ``a temperature (of something) is greater than another temperature''.}~\label{fig:parsing_structure_figure}
	\vspace{-0.7cm}
\end{figure*}

\section{PUMICE}
Motivated by the formative study results, we designed the \textsc{Pumice} agent that supports understanding ambiguous natural language instructions for task automation by allowing users to recursively define any new, ambiguous or vague concepts in a multi-level top-down process. 


\subsection{Example Scenario}
This section shows an example scenario to illustrate how \textsc{Pumice} works. Suppose a user starts teaching the agent a new task automation rule by saying, ``\textit{If it's hot, order a cup of Iced Cappuccino.}'' We also assume that the agent has no prior knowledge about the relevant task domains (weather and coffee ordering). Due to the lack of domain knowledge, the agent does not understand ``it's hot'' and ``order a cup of Iced Cappuccino''. However, the agent can recognize the conditional structure in the utterance (the parse for Utterance 0 in Figure~\ref{fig:parsing_structure_figure}) and can identify that ``it's hot'' should represent a Boolean expression while ``order a cup of Iced Cappuccino'' represents the action to perform if the condition is true.

\textsc{Pumice}'s semantic parser can mark unknown parts in user utterances using typed \texttt{resolve...()} functions, as marked in the yellow highlights in the parse for Utterance 0 in Figure~\ref{fig:parsing_structure_figure}. The \textsc{Pumice} agent then proceeds to ask the user to further explain these concepts. It asks, ``\textit{How do I tell whether it's hot?}'' since it has already figured out that ``it's hot'' should be a function that returns a Boolean value. The user answers ``\textit{It is hot when the temperature is above 85 degrees Fahrenheit.}'', as shown in Utterance 2 in Figure~\ref{fig:parsing_structure_figure}. \textsc{Pumice} understands the comparison (as shown in the parse for Utterance 2 in Figure~\ref{fig:parsing_structure_figure}), but does not know the concept of ``temperature'', only knowing that it should be a numeric value comparable to 85 degrees Fahrenheit. Hence it asks, ``\textit{How do I find out the value for temperature?}'', to which the user responds, ``\textit{Let me demonstrate for you.}''

Here the user can demonstrate the procedure of finding the current temperature by opening the weather app on the phone, and pointing at the current reading of the weather. To assist the user, \textsc{Pumice} uses a visualization overlay to highlight any GUI objects on the screen that fit into the comparison (i.e., those that display a value comparable to 85 degrees Fahrenheit). The user can choose from these highlighted objects (see Figure~\ref{fig:value_demo} for an example). Through this demonstration, \textsc{Pumice} learns a reusable procedure \texttt{query\textunderscore Temperature()} for getting the current value for the new concept \textit{temperature}, and stores it in a persistent knowledge graph so that it can be used in other tasks. \textsc{Pumice} confirms with the user every time it learns a new concept or a new rule, so that the user is aware of the current state of the system, and can correct any errors (see the Error Recovery and Backtracking section for details).

For the next phase, \textsc{Pumice} has already determined that ``order a cup of Iced Cappuccino'' should be an action triggered when the condition ``it's hot'' is true, but does not know how to perform this action (also known as intent fulfillment in chatbots~\cite{li_kite:_2018}). To learn how to perform this action, it asks, ``\textit{How do I order a cup of Iced Cappuccino?}'', to which the user responds, ``\textit{I can demonstrate.}'' The user then proceeds to demonstrate the procedure of ordering a cup of Iced Cappuccino using the existing app for Starbucks (a coffee chain). From the user demonstration, \textsc{Pumice} can figure out that ``Iced Cappuccino'' is a task parameter, and can learn the generalized procedure \texttt{order\textunderscore Starbucks()} for ordering any item in the Starbucks app, as well as a list of available items to order in the Starbucks app by looking through the Starbucks app's menus, using the underlying \textsc{Sugilite} framework [9] for processing the task recording.

Finally, \textsc{Pumice} asks the user about the else condition by saying, ``\textit{What should I do if it's not hot?}'' Suppose the user says ``\textit{Order a cup of Hot Latte},'' then the user will not need to demonstrate again because \textsc{Pumice} can recognize ``hot latte'' as an available parameter for the known \texttt{order\textunderscore Starbucks()} procedure. 

\subsection{Design Features}
In this section, we discuss several of \textsc{Pumice}'s key design features in its user interactions, and how they were motivated by results of the formative study.

\subsubsection{Support for Concept Learning}
In the formative study, we identified two main challenges in regards to concept learning. First, user often naturally use intrinsically unclear or ambiguous concepts when instructing intelligent agents (e.g., ``\textit{register for easy courses}'', where the definition of ``easy'' depends on the context and the user preference). Second, users expect agents to understand common-sense concepts that the agents may not know. To address these challenges, we designed and implemented the support for concept learning in \textsc{Pumice}. \textsc{Pumice} can detect and learn three kinds of unknown components in user utterances: \textit{procedures}, \textit{Boolean concepts}, and \textit{value concepts}. Because \textsc{Pumice}'s support for procedure learning is unchanged from the underlying \textsc{Sugilite} mechanisms~\cite{li_appinite:_2018,li_sugilite:_2017}, in this section, we focus on discussing how \textsc{Pumice} learns Boolean concepts and value concepts. 

When encountering an unknown or unclear concept in the utterance parsing result, \textsc{Pumice} first determines the concept type based on the context. If the concept is used as a condition (e.g., ``if \textbf{it is hot}''), then it should be of Boolean type. Similarly, if a concept is used where a value is expected (e.g., ``if the \textbf{current temperature} is above 70\textdegree F'' or ``set the AC to the \textbf{current temperature}''), then it will be marked as a value concept. Both kinds of concepts are represented as typed \texttt{resolve()} functions in the parsing result (shown in Figure~\ref{fig:parsing_structure_figure}), indicating that they need to be further resolved down the line. This process is flexible. For example, if the user clearly defines a condition without introducing unknown or unclear concepts, then \textsc{Pumice} will not need to ask follow-up questions for concept resolution.

\textsc{Pumice} recursively executes each \texttt{resolve()} function in the parsing result in a depth-first fashion. After a concept is fully resolved (i.e., all concepts used for defining it have been resolved), it is added to a persistent knowledge graph (details in the System Implementation section), and a link to it replaces the \texttt{resolve()} function. From the user's perspective, when a \texttt{resolve()} function is executed, the agent asks a question to prompt the user to further explain the concept. When resolving a Boolean concept, \textsc{Pumice} asks, ``\textit{How do I know whether [concept\textunderscore name]?}'' For resolving a value concept, \textsc{Pumice} asks, ``\textit{How do I find out the value of [concept\textunderscore name]?}''

To explain a new Boolean concept, the user may verbally refer to another Boolean concept (e.g., ``traffic is heavy'' means ``commute takes a long time'') or may describe a Boolean expression (e.g., ``the commute time is longer than 30 minutes''). When describing the Boolean expression, the user can use flexible words (e.g., colder, further, more expensive) to describe the relation (i.e., greater than, less than, and equal to). As explained previously, if any new Boolean or value concepts are used in the explanation, \textsc{Pumice} will recursively resolve them. The user can also use more than one unknown value concepts, such as ``if the price of a Uber is greater than the price of a Lyft'' (Uber and Lyft are both popular ridesharing apps).

Similar to Boolean concepts, the user can refer to another value concept when explaining a value concept. When a value concept is concrete and available in a mobile app, the user can also demonstrate how to query the value through app GUIs. The formative study has suggested that this multi-modal approach is effective and feasible for end users.  After users indicate that they want to demonstrate, \textsc{Pumice} switches to the home screen of the phone, and prompts the user to demonstrate how to find out the value of the concept. 

\begin{figure}
	\centering
	\includegraphics[width=\columnwidth]{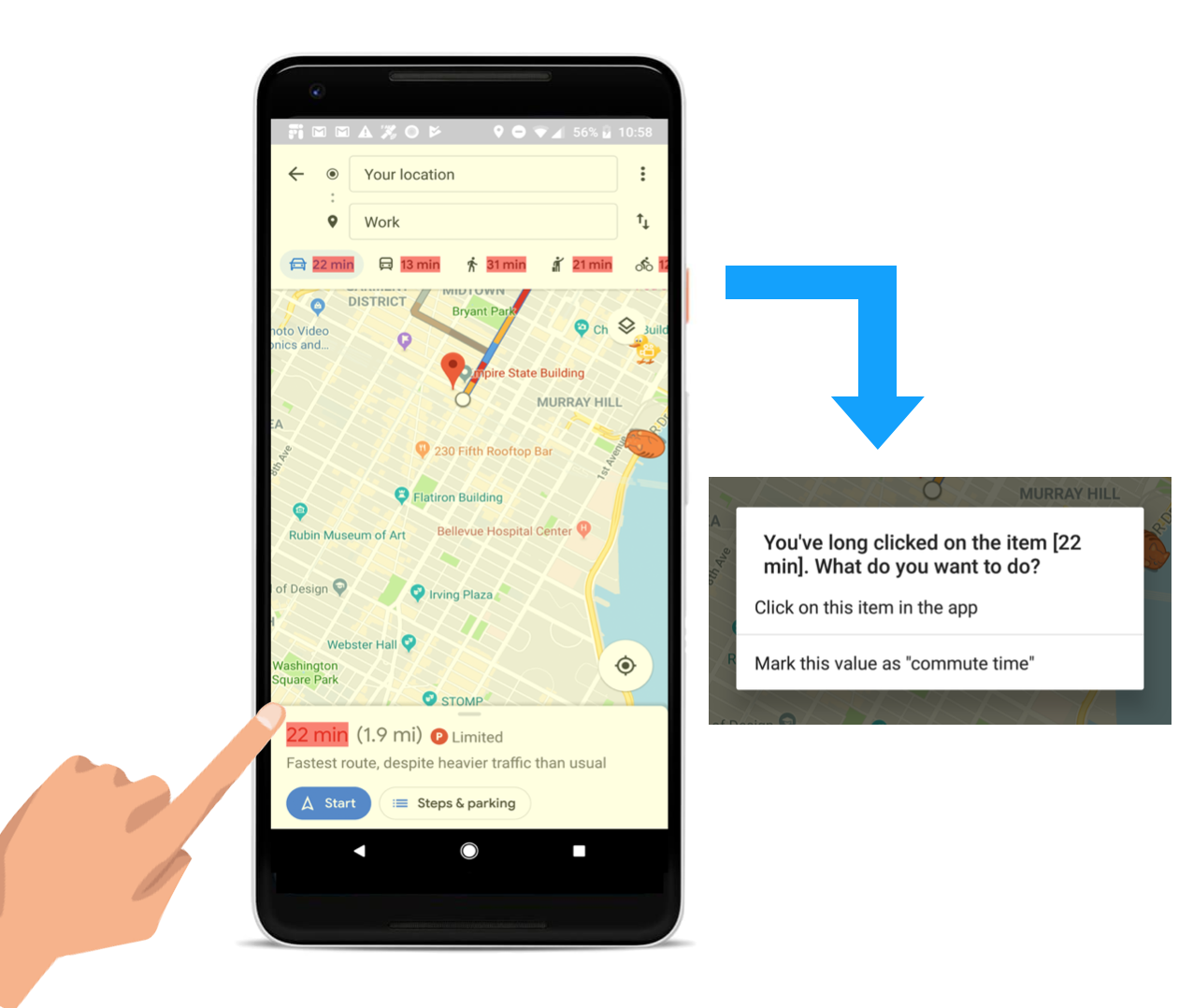}
	\vspace{-0.4cm}
	\caption{The user teaches the value concept ``commute time'' by demonstrating querying the value in Google Maps. The red overlays highlight all durations it was able to identify on the Google Maps GUI.}~\label{fig:value_demo}
	\vspace{-0.7cm}
\end{figure}

To help the user with value concept demonstrations, \textsc{Pumice} highlights possible items on the current app GUI if the type of the target concept can be inferred from the type of the constant value, or using the type of value concept to which it is being compared (see Figure~\ref{fig:value_demo}). For example, in the aforementioned ``commute time'' example, \textsc{Pumice} knows that ``commute time'' should be a duration, because it is comparable to the constant value ``30 minutes''. Once the user finds the target value in an app, they can long press on the target value to select it and indicate it as the target value. \textsc{Pumice} uses an interaction proxy overlay~\cite{zhang_interaction_2017} for recording, so that it can record \textit{all} values visible on the screen, not limited to the selectable or clickable ones. \textsc{Pumice} can extract these values from the GUI using the screen scraping mechanism in the underlying \textsc{Sugilite} framework~\cite{li_sugilite:_2017}. Once the target value is selected, \textsc{Pumice} stores the procedure of navigating to the screen where the target value is displayed and finding the target value on the screen into its persistent knowledge graph as a value query, so that this query can be used whenever the underlying value is needed. After the value concept demonstration, \textsc{Pumice} switches back to the conversational interface and continues to resolve the next concept if needed.

\subsubsection{Concept Generalization and Reuse}
Once concepts are learned, another major challenge is to generalize them so that they can be reused correctly in different contexts and task domains. This is a key design goal of \textsc{Pumice}. It should be able to learn concepts at a fine granularity, and reuse parts of existing concepts as much as possible to avoid asking users to make redundant demonstrations. In our previous works on generalization for PBD, we focused on generalizing procedures, specifically learning parameters~\cite{li_sugilite:_2017} and intents for underlying operations~\cite{li_appinite:_2018}. We have already deployed these existing generalization mechanisms in \textsc{Pumice}, but in addition, we also explored the generalization of Boolean concepts and value concepts.  

When generalizing Boolean concepts, \textsc{Pumice} assumes that the Boolean operation stays the same, but the arguments may differ. For example, for the concept ``hot'' in Figure~\ref{fig:parsing_structure_figure}, it should still mean that a temperature (of something) is greater than another temperature. But the two  in comparison can be different constants, or from different value queries. For example, suppose after the interactions in Figure~\ref{fig:parsing_structure_figure}, the user instructs a new rule ``\textit{if the oven is hot, start the cook timer.}'' \textsc{Pumice} can recognize that ``hot'' is a concept that has been instructed before in a different context, so it asks ``\textit{I already know how to tell whether it is hot when determining whether to order a cup of Iced Cappuccino. Is it the same here when determining whether to start the cook timer?}'' After responding ``No'', the user can instruct how to find out the temperature of the oven, and the new threshold value for ``hot'' either by instructing a new value concept, or using a constant value.

The generalization mechanism for value concepts works similarly. \textsc{Pumice} supports value concepts that share the same name to have different query implementations for different task contexts. For example, following the ``if the oven is hot, start the cook timer'' example, suppose the user defines ``hot'' for this new context as ``\textit{The temperature is above 400 degrees.}'' \textsc{Pumice} realizes that there is already a value concept named ``temperature'', so it will ask ``\textit{I already know how to find out the value for temperature using the Weather app. Should I use that for determining whether the oven is hot?}'', to which the user can say ``No'' and then demonstrate querying the temperature of the oven using the corresponding app (assuming the user has a smart oven with an in-app display of its temperature).

This mechanism allows concepts like ``hot'' to be reused at three different levels: (1) exactly the same (e.g., the temperature of the weather is greater than 85\textdegree F); (2) different threshold (e.g., the temperature of the weather is greater than \textit{x}); and (3) different value query (e.g., the temperature of \textit{something else} is greater than \textit{x}).

\subsubsection{Error Recovery and Backtracking}
Like all other interactive EUD systems, it is crucial for \textsc{Pumice} to properly handle errors, and to backtrack from errors in speech recognition, semantic parsing, generalizations, and inferences of intent~\cite{li_teaching_2018}. We iteratively tested early prototypes of \textsc{Pumice} with users through early usability testing, and developed the following mechanisms to support error recovery and backtracking in \textsc{Pumice}.  

To mitigate semantic parsing errors, we implemented a mixed-initiative mechanism where \textsc{Pumice} can ask users about \textit{components} within the parsed expression if the parsing result is considered incorrect by the user. Because parsing result candidates are all typed expressions in \textsc{Pumice}'s internal functional domain-specific language (DSL) as a conditional, Boolean, value, or procedure, \textsc{Pumice} can identify components in a parsing result that it is less confident about by comparing the top candidate with the alternatives and confidence scores, and ask the user about them.

For example, suppose the user defines a Boolean concept ``good restaurant'' with the utterance ``the rating is better than 2''. The parser is uncertain about the comparison operator in this Boolean expression, since ``better'' can mean either ``greater than'' or ``less than'' depending on the context. It will ask the user ``\textit{I understand you are trying to compare the value concept `rating' and the value `2', should `rating' be greater than, or less than `2'?}'' The same technique can also be used to disambiguate other parts of the parsing results, such as the argument of \texttt{resolve()} functions (e.g., determining whether the unknown procedure should be ``order a cup of Iced Cappuccino'' or ``order a cup''  for Utterance 0 in Figure~\ref{fig:parsing_structure_figure}).

\textsc{Pumice} also provides an ``undo'' function to allow the user to backtrack to a previous conversational state in case of incorrect speech recognition, incorrect generalization, or when the user wants to modify a previous input. Users can either say that they want to go back to the previous state, or click on an ``undo'' option in \textsc{Pumice}'s menu (activated from the option icon on the top right corner on the screen shown in Figure~\ref{fig:parsing_structure_figure}).

\subsection{System Implementation}
We implemented the \textsc{Pumice} agent as an Android app. The app was developed and tested on a Google Pixel 2 XL phone running Android 8.0. \textsc{Pumice} does \textit{not} require the root access to the phone, and should run on any phone running Android 6.0 or higher. \textsc{Pumice} is open-sourced on GitHub\footnote{https://github.com/tobyli/Sugilite\_development}. 

\subsubsection{Semantic Parsing}
We built the semantic parser for \textsc{Pumice} using the \textsc{Sempre} framework~\cite{berant_semantic_2013}. The parser runs on a remote Linux server, and communicates with the \textsc{Pumice} client through an HTTP RESTful interface. It uses the Floating Parser architecture, which is a grammar-based approach that provides more flexibility without requiring hand-engineering of lexicalized rules like synchronous CFG or CCG based semantic parsers~\cite{pasupat_compositional_2015}. This approach also provides more interpretable results and requires less training data than neural network approaches (e.g.,~\cite{Yin:8595231,DBLP:journals/corr/YinN17}). The parser parses user utterances into expressions in a simple functional DSL we created for \textsc{Pumice}.

A key feature we added to \textsc{Pumice}'s parser is allowing typed \texttt{resolve()} functions in the parsing results to indicate unknown or unclear concepts and procedures. This feature adds interactivity to the traditional semantic parsing process. When this \texttt{resolve()} function is called at runtime, the front-end \textsc{Pumice} agent asks the user to verbally explain, or to demonstrate how to fulfill this \texttt{resolve()} function. If an unknown concept or procedure is resolved through verbal explanation, the parser can parse the new explanation into an expression of its original type in the target DSL (e.g., an explanation for a Boolean concept is parsed into a Boolean expression), and replace the original \texttt{resolve()} function with the new expression. The parser also adds relevant utterances for existing concepts and procedures, and visible text labels from demonstrations on third-party app GUIs to its set of lexicons, so that it can understand user references to those existing knowledge and in-app contents. \textsc{Pumice}'s parser was trained on rich features that associate lexical and syntactic patterns (e.g., unigrams, bigrams, skipgrams, part-of-speech tags, named entity tags) of user utterances with semantics and structures of the target DSL over a small number of training data ($n=905$) that were mostly collected and enriched from the formative study.\looseness=-1

\subsubsection{Demonstration Recording and Replaying}
\textsc{Pumice} uses our open-sourced \textsc{Sugilite}~\cite{li_sugilite:_2017} framework to support its demonstration recording and replaying. \textsc{Sugilite} provides action recording and replaying capabilities on third-party Android apps using the Android Accessibility API. \textsc{Sugilite} also provides the support for parameterization of sequences of actions (e.g., identifying ``Iced Cappuccino'' as a parameter and ``Hot Latte'' as an alternative value in the example in Figure~\ref{fig:parsing_structure_figure}), and the support for handling minor GUI changes in apps. Through \textsc{Sugilite}, \textsc{Pumice} operates well on most native Android apps, but may have problems working with web apps and apps with special graphic engines (e.g., games). It currently does not support recording gestural and sensory inputs (e.g., rotating the phone) either.

\subsubsection{Knowledge Representations}
\textsc{Pumice} maintains two kinds of knowledge representations: a continuously refreshing UI snapshot graph representing third-party app contexts for demonstration, and a persistent knowledge base for storing learned procedures and concepts. 

\begin{figure}
	\centering
	\includegraphics[width=\columnwidth]{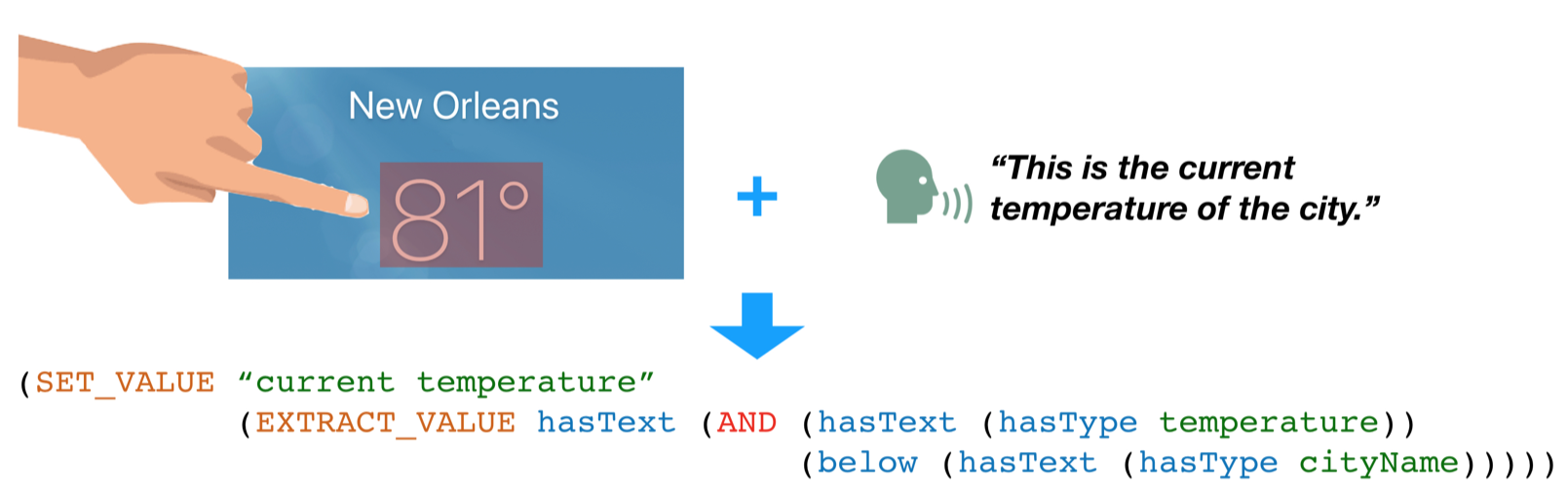}
	\caption{An example showing how \textsc{Pumice} parses the user's demonstrated action and verbal reference to an app's GUI content into a \texttt{SET\textunderscore VALUE} statement with a query over the UI snapshot graph when resolving a new value concept ``current temperature''}~\label{fig:ui_parsing_example}
	\vspace{-0.7cm}
\end{figure}

The purpose of the UI snapshot graph is to support understanding the user's references to app GUI contents in their verbal instructions. The UI snapshot graph mechanism used in \textsc{Pumice} was extended from \textsc{Appinite}~\cite{li_appinite:_2018}. For every state of an app's GUI, a UI snapshot graph is constructed to represent \textit{all} visible and invisible GUI objects on the screen, including their types, positions, accessibility labels, text labels, various properties, and spatial relations among them. We used a lightweight semantic parser from the Stanford CoreNLP~\cite{manning_stanford_2014} to extract types of structured data (e.g., temperature, time, date, phone number) and named entities (e.g., city names, people's names). When handling the user's references to app GUI contents, \textsc{Pumice} parses the original utterances into queries over the current UI snapshot graph (example in Figure~\ref{fig:ui_parsing_example}). This approach allows \textsc{Pumice} to generate flexible queries for value concepts and procedures that accurately reflect user intents, and which can be reliably executed in future contexts. 

The persistent knowledge base stores all procedures, concepts, and facts \textsc{Pumice} has learned from the user. Procedures are stored as \textsc{Sugilite}~\cite{li_sugilite:_2017} scripts, with the corresponding trigger utterances, parameters, and possible alternatives for each parameter. Each Boolean concept is represented as a set of trigger utterances, Boolean expressions with references to the value concepts or constants involved, and contexts (i.e., the apps and actions used) for each Boolean expression. Similarly, the structure for each stored value concept includes its triggering utterances, demonstrated value queries for extracting target values from app GUIs, and contexts for each value query. 


\section{User Study}
We conducted a lab study to evaluate the usability of \textsc{Pumice}. In each session, a user completed 4 tasks. For each task, the user instructed \textsc{Pumice} to create a new task automation, with the required conditionals and new concepts. We used a task-based method to specifically test the usability of \textsc{Pumice}'s design, since the motivation for the design derives from the formative study results. We did not use a control condition, as we could not find other tools that can feasibly support users with little programming expertise to complete the target tasks. 

\subsection{Participants}
We recruited 10 participants (5 women, 5 men, ages 19 to 35) for our study. Each study session lasted 40 to 60 minutes. We compensated each participant \$15 for their time. 6 participants were students in two local universities, and the other 4 worked different technical, administrative or managerial jobs. All participants were experienced smartphone users who had been using smartphones for at least 3 years. 8 out of 10 participants had some prior experience of interacting with conversational agents like Siri, Alexa and Google Assistant.

We asked the participants to report their programming experience on a five-point scale from ``never programmed'' to ``experienced programmer''. Among our participants, there were 1 who had never programmed, 5 who had only used end-user programming tools (e.g., Excel functions, Office macros), 1 novice programmer with experience equivalent to 1-2 college level programming classes, 1 programmer with 1-2 years of experience, and 2 programmers with more than 3 years of experience. In our analysis, we will label the first two groups ``non-programmers'' and the last three groups ``programmers''.

\subsection{Procedure}
At the beginning of each session, the participant received a 5-minute tutorial on how to use \textsc{Pumice}. In the tutorial, the experimenter demonstrated an example of teaching \textsc{Pumice} to check the bus schedule when ``it is late'', and ``late'' was defined as ``current time is after 8pm'' through a conversation with \textsc{Pumice}. The experimenter then showed how to demonstrate to \textsc{Pumice} finding out the current time using the Clock app. 

Following the tutorial, the participant was provided a Google Pixel 2 phone with \textsc{Pumice} and relevant third-party apps installed. The experimenter showed the participant the available apps, and made sure that the participant understood the functionality of each third-party app. We did this because the underlying assumption of the study (and the design of \textsc{Pumice}) is that users are familiar with the third-party apps, so we are testing whether they can successfully use \textsc{Pumice}, not the apps. Then, the participant received 4 tasks in random order. We asked participants to keep trying until they were able to correctly execute the automation, and that they were happy with the resulting actions of the agent. We also checked the scripts at the end of each study session to evaluate their correctness. 

After completing the tasks, the participant filled out a post-survey to report the perceived usefulness, ease of use and naturalness of interactions with \textsc{Pumice}. We ended each session with a short informal interview with the participant on their experiences with \textsc{Pumice}.

\subsection{Tasks}
We assigned 4 tasks to each participant. These tasks were designed by combining common themes observed in users' proposed scenarios from the formative study. We ensured that these tasks (1) covered key \textsc{Pumice} features (i.e., concept learning, value query demonstration, procedure demonstration, concept generalization, procedure generalization and ``else'' condition handling); (2) involved only app interfaces that most users are familiar with; and (3) used conditions that we can control so we can test the correctness of the scripts (we controlled the temperature, the traffic condition, and the room price by manipulating the GPS location of the phone).\looseness=-1

\begin{figure}
	\centering
	\includegraphics[width=0.8\columnwidth]{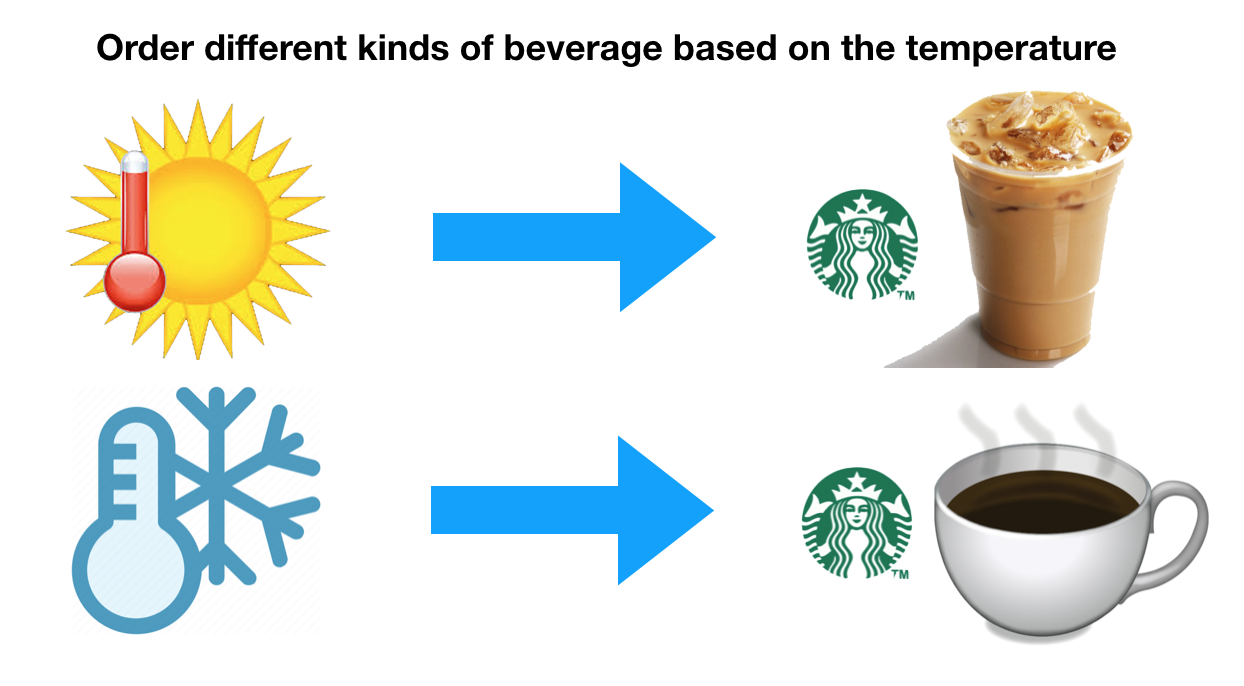}
	\caption{The graphical prompt used for Task 1 -- A possible user command can be ``\textit{Order Iced coffee when it's hot outside, otherwise order hot coffee when the weather is cold.}'' }~\label{fig:study_graphic_example}
	\vspace{-0.8cm}
\end{figure}

In order to minimize biasing users' utterances, we used the Natural Programming Elicitation method~\cite{myers_programmers_2016}. Task descriptions were provided in the form of graphics, with minimal text descriptions that could not be directly used in user instructions (see Figure~\ref{fig:study_graphic_example} for an example). 

\paragraph{Task 1}
In this task, the user instructs \textsc{Pumice} to order iced coffee when the weather is hot, and order hot coffee otherwise (Figure~\ref{fig:study_graphic_example}). We pre-taught \textsc{Pumice} the concept of ``hot'' in the task domain of turning on the air conditioner. So the user needs to utilize the concept generalization feature to generalize the existing concept ``hot'' to the new domain of coffee ordering. The user also needs to demonstrate ordering iced coffee using the Starbucks app, and to provide ``order hot coffee'' as the alternative for the ``else'' operation. The user does not need to demonstrate again for ordering hot coffee, as it can be automatically generalized from the previous demonstration of ordering iced coffee. 

\paragraph{Task 2}
In this task, the user instructs \textsc{Pumice} to set an alarm for 7:00am if the traffic is heavy on their commuting route. We pre-stored ``home'' and ``work'' locations in Google Maps. The user needs to define ``heavy traffic'' as prompted by \textsc{Pumice} by demonstrating how to find out the estimated commute time, and explaining that ``heavy traffic'' means that the commute takes more than 30 minutes. The user then needs to demonstrate setting a 7:00am alarm using the built-in Clock app. 

\paragraph{Task 3}
In this task, the user instructs \textsc{Pumice} to choose between making a hotel reservation and requesting a Uber to go home depending on whether the hotel price is cheap. The user should verbally define "cheap" as "room price is below \$100", and demonstrate how to find out the hotel price using the Marriott (a hotel chain) app. The user also needs to demonstrate making the hotel reservation using the Marriott app, specify "request an Uber" as the action for the ``else'' condition, and demonstrate how to request an Uber using the Uber app.

\paragraph{Task 4}
In this task, the user instructs \textsc{Pumice} to order a pepperoni pizza if there is enough money left in the food budget. The user needs to define the concept of ``enough budget'', demonstrate finding out the balance of the budget  using the Spending Tracker app, and demonstrate ordering a pepperoni pizza using the Papa Johns (a pizza chain) app. 

\begin{figure}
	\centering
	\includegraphics[width=\columnwidth]{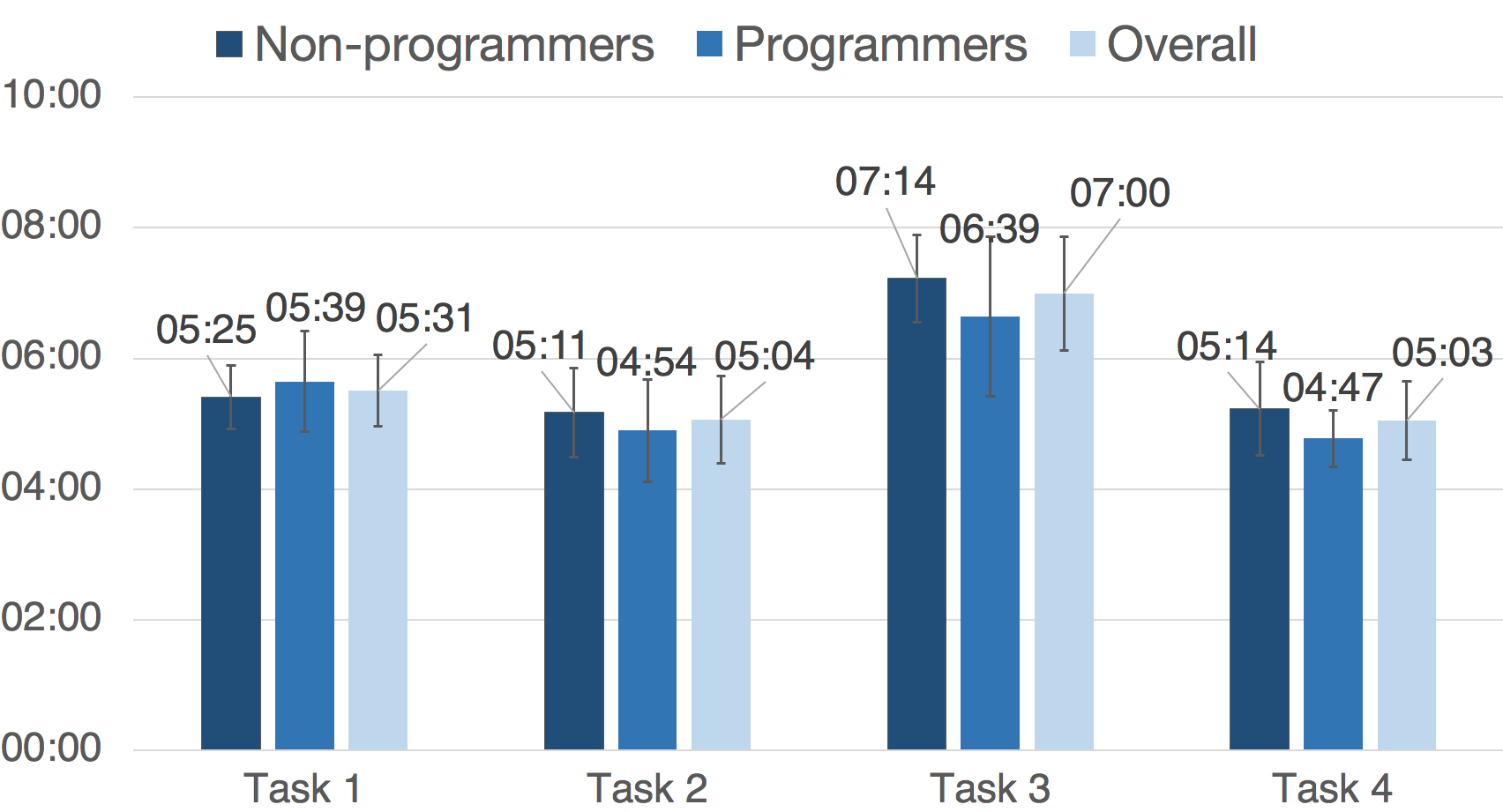}
	\vspace{-0.3cm}
	\caption{The average task completion times for each task. The error bars show one standard deviation in each direction.  }~\label{fig:study_result}
	\vspace{-1cm}
\end{figure}

\subsection{Results}
All participants were able to complete all 4 tasks. The total time for tasks ranged from 19.4 minutes to 25 minutes for the 10 participants. Figure~\ref{fig:study_result} shows the overall average task completion time of each task, as well as the comparison between the non-programmers and the programmers. The average total time-on-task for programmers (22.12 minutes, SD=2.40) was slightly shorter than that for non-programmers (23.06 minutes, SD=1.57), but the difference was not statistically significant.

Most of the issues encountered by participants were actually from the Google Cloud speech recognition system used in \textsc{Pumice}. It would sometimes misrecognize the user's voice input, or cut off the user early. These errors were handled by the participants using the ``undo'' feature in \textsc{Pumice}. Some participants also had parsing errors. \textsc{Pumice}'s current semantic parser has limited capabilities in understanding references of pronouns (e.g., for an utterance \textit{``it takes longer than 30 minutes to get to work''}, the parser would recognize it as ``it'' instead of ``the time it takes to get to work'' is greater than 30 minutes). Those errors were also handled by participants through undoing and rephrasing. One participant ran into the ``confusion of Boolean operator'' problem in Task 2 when she used the phrase ``\textit{commute [time is] worse than 30 minutes}'', for which the parser initially recognized incorrectly as ``commute is \textit{less than} 30 minutes.'' She was able to correct this using the mixed-initiative mechanism, as described in the Error Recovery and Backtracking section.

Overall, no participant had major problem with the multi-modal interaction approach and the top-down recursive concept resolution conversational structure, which was encouraging. However, all participants had received a tutorial with an example task demonstrated. We also emphasized in the tutorial that they should try to use concepts that can be found in mobile apps in their explanations of new concepts. These factors might contributed to the successes of our participants.\looseness=-1

In a post survey, we asked participants to rate statements about the usability, naturalness and usefulness of \textsc{Pumice} on a 7-point Likert scale from ``strongly disagree'' to ``strongly agree''. \textsc{Pumice} scored on average $6.2$ on ``\textit{I feel \textsc{Pumice} is easy to use}'', $6.1$ on ``\textit{I find my interactions with \textsc{Pumice} natural}'', and $6.9$ on ``\textit{I think \textsc{Pumice} is a useful tool for automating tasks on smartphones,}'' indicating that our participants were generally satisfied with their experience using \textsc{Pumice}.

\subsection{Discussion}
In the informal interview after completing the tasks, participants praised \textsc{Pumice} for its naturalness and low learning barriers. Non-programmers were particularly impressed by the multi-modal interface. For example, P7 (who was a non-programmer) said: ``\textit{Teaching \textsc{Pumice} feels similar to explaining tasks to another person...[\textsl{Pumice}'s] support for demonstration is very easy to use since I'm already familiar with how to use those apps.}'' Participants also considered \textsc{Pumice}'s top-down interactive concept resolution approach very useful, as it does not require them to define everything clearly upfront. 

Participants were excited about the usefulness of \textsc{Pumice}. P6 said, ``\textit{I have an Alexa assistant at home, but I only use them for tasks like playing music and setting alarms. I tried some more complicated commands before, but they all failed. If it had the capability of \textsc{Pumice}, I would definitely use it to teach Alexa more tasks.}'' They also proposed many usage scenarios based on their own needs in the interview, such as paying off credit card balance early when it has reached a certain amount, automatically closing background apps when the available phone memory is low, monitoring promotions for gifts saved in the wish list when approaching anniversaries, and setting reminders for events in mobile games.   

Several concerns were also raised by our participants. P4 commented that \textsc{Pumice} should ``just know'' how to find out weather conditions without requiring her to teach it since ``all other bots know how to do it'', indicating the need for a hybrid approach that combines EUD with pre-programmed common functionalities. P5 said that teaching the agent could be too time-consuming unless the task was very repetitive since he could just ``do it with 5 taps.'' Several users also expressed privacy concerns after learning that \textsc{Pumice} can see all screen contents during demonstrations, while one user, on the other hand, suggested having \textsc{Pumice} observe him at all times so that it can learn things in the background.

\section{Limitations and Future Work}
The current version of \textsc{Pumice} has no semantic understanding of information involved in tasks, which prevents it from dealing with implicit parameters (e.g., ``when it snows'' means ``the current weather condition is \textit{snowing}'') and understanding the relations between concepts (e.g., Iced Cappuccino and Hot Latte are both instances of \textit{coffee}; Iced Cappuccino has the property of being \textit{cold}). The parser also does not process references, synonyms, antonyms, or implicit conjunctions/disjunctions in utterances. We plan to address these problems by leveraging more advanced NLP techniques. Specifically, we are currently investigating bringing in external sources of world knowledge (e.g., Wikipedia, Freebase~\cite{bollacker_freebase:_2008}, ConceptNet~\cite{Liu2004}, WikiBrain~\cite{sen_wikibrain:_2014}, or NELL~\cite{mitchell2018never}), which can enable more intelligent generalizations, suggestions, and error detection. The agent can also make better guesses when dealing with ambiguous user inputs. As discussed previously, \textsc{Pumice} already uses relational structures to store the context of app GUIs and its learned knowledge, which should make it easier to incorporate external knowledge graphs.

In the future, we plan to expand \textsc{Pumice}'s expressiveness in representing conditionals and Boolean expressions. In the current version, it only supports single basic Boolean operations (i.e., greater than, less than, equal to) without support for logical operations (e.g., when the weather is cold \textit{and} raining) or arithmetic operations (e.g., if is at least \textit{\$10 more expensive} than Lyft), or counting GUI elements (e.g., ``highly rated'' means \textit{more than 3 stars are red}) We plan to explore the design space of new interactive interfaces to support these features in future versions. Note that it will likely require more than just adding grammar rules to the semantic parser and expanding the DSL, since end users' usage of words like ``and'' and ``or'', and their language for specifying computation are known to often be ambiguous~\cite{pane_studying_2001}.

Further, although \textsc{Pumice} supports generalization of procedures, Boolean concepts and value concepts across different task domains, all such generalizations are stored locally on the phone and limited to one user. We plan to expand \textsc{Pumice} to support generalizing learned knowledge across multiple users. The current version of \textsc{Pumice} does not differentiate between personal preferences and generalizable knowledge in learned concepts and procedures. An important focus of our future work is to distinguish these, and allow the sharing and aggregation of generalizable components across multiple users. To support this, we will also need to develop appropriate mechanisms to help preserve user privacy.

In this prototype of \textsc{Pumice}, the proposed technique is used in conversations for performing immediate tasks instead of for completely automated rules. We plan to add the support for automated rules in the future. An implementation challenge for supporting automated rules is to continuously poll values from GUIs. The current version of underlying \textsc{Sugilite} framework can only support foreground execution, which is not feasible for background monitoring for triggers. We plan to use techniques like virtual machine (VM) to support background execution of demonstrated scripts.

Lastly, we plan to conduct an open-ended field study to better understand how users use \textsc{Pumice} in real-life scenarios. Although the design of \textsc{Pumice} was motivated from results of a formative study with users, and the usability of \textsc{Pumice} has been supported by an in-lab user study, we hope to further understand what tasks users choose to program, how they switch between different input modalities, and how useful \textsc{Pumice} is for users in realistic contexts.

\section{Conclusion}
We have presented \textsc{Pumice}, an agent that can learn concepts and conditionals from conversational natural language instructions and demonstrations. Through \textsc{Pumice}, we showcased the idea of using multi-modal interactions to support the learning of unknown, ambiguous or vague concepts in users' verbal commands, which were shown to be common in our formative study. 

In \textsc{Pumice}'s approach, users can explain abstract concepts in task conditions using more concrete smaller concepts, and ground them by demonstrating with third-party mobile apps. More broadly, our work demonstrates how combining conversational interfaces and demonstrational interfaces can create easy-to-use and natural end user development experiences.

\section{Acknowledgments}
This research was supported in part by Verizon and Oath through the InMind project, a J.P. Morgan Faculty Research Award, and NSF grant IIS-1814472. Any opinions, findings and conclusions or recommendations expressed in this material are those of the authors and do not necessarily reflect the views of the sponsors. We would like to thank our study participants for their help, our anonymous reviewers, and Michael Liu, Fanglin Chen, Haojian Jin, Brandon Canfield, Jingya Chen, and William Timkey for their insightful feedback.\looseness=-1

\bibliographystyle{aaai}
\bibliography{references_2}

\end{document}